\documentclass[letterpaper]{article} % DO NOT CHANGE THIS
\usepackage{aaai25}  % DO NOT CHANGE THIS
\usepackage{times}  % DO NOT CHANGE THIS
\usepackage{helvet}  % DO NOT CHANGE THIS
\usepackage{courier}  % DO NOT CHANGE THIS
\usepackage[hyphens]{url}  % DO NOT CHANGE THIS
\usepackage{graphicx} % DO NOT CHANGE THIS
\usepackage{xcolor}
\usepackage{afterpage}
\usepackage{multicol}
\usepackage{longtable}
\usepackage{booktabs}
\usepackage{lscape} % Para paisagem, caso seja necessário

\usepackage{natbib}  % DO NOT CHANGE THIS AND DO NOT ADD ANY OPTIONS TO IT
\usepackage{caption} % DO NOT CHANGE THIS AND DO NOT ADD ANY OPTIONS TO IT
\usepackage{aaai25}  % DO NOT CHANGE THIS
\usepackage{times}  % DO NOT CHANGE THIS
\usepackage{helvet}  % DO NOT CHANGE THIS
\usepackage{courier}  % DO NOT CHANGE THIS
\usepackage[hyphens]{url}  % DO NOT CHANGE THIS
\usepackage{graphicx} % DO NOT CHANGE THIS
\usepackage{natbib}  % DO NOT CHANGE THIS AND DO NOT ADD ANY OPTIONS TO IT
\usepackage{caption} % DO NOT CHANGE THIS AND DO NOT ADD ANY OPTIONS TO IT
\DeclareCaptionStyle{ruled}{labelfont=normalfont,labelsep=colon,strut=off} % DO NOT CHANGE THIS
\usepackage{algorithm}
\usepackage{algorithmic}
\usepackage{xcolor}
\newcommand{\answerYes}[1]{\textcolor{blue}{#1}} 
 
\newcommand{\answerNA}[1]{\textcolor{gray}{#1}} 
 
\usepackage{algorithm}
\usepackage{algorithmic}

\usepackage{newfloat}
\usepackage{listings}
\DeclareCaptionStyle{ruled}{labelfont=normalfont,labelsep=colon,strut=off} % DO NOT CHANGE THIS
\floatstyle{ruled}
\newfloat{listing}{tb}{lst}{}
\floatname{listing}{Listing}
\nocopyright % -- Your paper will not be published if you use this command
\title{Discord Unveiled: A Comprehensive Dataset of Public Communication (2015-2024)}

\author{
    Yan Aquino \textsuperscript{\rm 1},
    Pedro Bento \textsuperscript{\rm 1}, 
    Arthur Buzelin \textsuperscript{\rm 1},
    Lucas Dayrell \textsuperscript{\rm 1}, 
    Samira Malaquias \textsuperscript{\rm 1}, \\
    Caio Santana \textsuperscript{\rm 1}, 
    Victoria Estanislau \textsuperscript{\rm 1},
    Pedro Dutenhefner \textsuperscript{\rm 1},
    Guilherme H. G. Evangelista \textsuperscript{\rm 1},\\
    Luisa G. Porfírio \textsuperscript{\rm 1},
    Caio Souza Grossi \textsuperscript{\rm 1},
    Pedro B. Rigueira \textsuperscript{\rm 1},\\
    Virgilio Almeida \textsuperscript{\rm 1},
    Gisele L. Pappa \textsuperscript{\rm 1},
    Wagner Meira Jr. \textsuperscript{\rm 1}
}
\affiliations{
    \textsuperscript{\rm 1}Universidade Federal de Minas Gerais - UFMG\\
    Belo Horizonte, Brazil\\
    \{yanaquino, pedro.bento, arthurbuzelin, lucasdayrell, samiramalaquias, caiosantana, victoria.estanislau, guilherme.evangelhista, luisagontijo, caio.grossi, pedrobacelar.rigueira, virgilio, glpappa, meira\}@dcc.ufmg.br, pedroroblesduten@ufmg.br
    
}

\usepackage{bibentry}
\begin{document}

\maketitle

\begin{abstract}
Discord has evolved from a gaming-focused communication tool into a versatile platform supporting diverse online communities. Despite its large user base and active public servers, academic research on Discord remains limited due to data accessibility challenges. This paper introduces \textbf{Discord Unveiled: A Comprehensive Dataset of Public Communication (2015-2024)}, the most extensive Discord public server's data to date. The dataset comprises over \textbf{2.05 billion messages} from \textbf{4.74 million users} across \textbf{3,167 public servers}, representing approximately 10\% of servers listed in Discord’s Discovery feature. Spanning from Discord’s launch in 2015 to the end of 2024, it offers a robust temporal and thematic framework for analyzing decentralized moderation, community governance, information dissemination, and social dynamics. Data was collected through Discord’s public API, adhering to ethical guidelines and privacy standards via anonymization techniques. Organized into structured JSON files, the dataset facilitates seamless integration with computational social science methodologies. Preliminary analyses reveal significant trends in user engagement, bot utilization, and linguistic diversity, with English predominating alongside substantial representations of Spanish, French, and Portuguese. Additionally, prevalent community themes such as social, art, music, and memes highlight Discord’s expansion beyond its gaming origins.

\end{abstract}

\section{Introduction}

Over the past decades, platforms such as Facebook, Instagram, and Twitter have played pivotal roles in online social interactions, significantly influencing political campaigns, information dissemination, digital community building, and shaping cultural dynamics. Previous studies have underscored the necessity of analyzing social platforms to understand social phenomena \cite{hate-speech-reddit-savvas, moderation-importance-qanon, facebook-1}. In this context, Discord emerges as a unique platform due to its decentralized nature, where moderation is explicitly delegated to users themselves\footnote{https://discord.com/community/your-responsibilities-as-a-discord-moderator-discord}. This model creates a potentially volatile environment for exposing social and cultural phenomena that might be overlooked on more centralized platforms. As other social networks have recently begun adopting trends of user-driven moderation, such as the use of community notes for fact-checking, Discord, with its longstanding user-moderation approach, offers valuable insights into what the future of decentralized moderation systems might evolve into.

The study of the phenomena generated by these online interactions was only possible due to data availability, and the emergence of new research areas interested in them opened up discussions on the pros and cons of the wide adoption of online social media. However, access to data from these platforms has become increasingly restricted. Recently, Twitter has severely limited its public Application Programming Interface (API)\footnote{https://docs.x.com/x-api/introduction}, while Meta (the parent company of Facebook and Instagram) also tightened data availability through its APIs\footnote{https://developers.facebook.com/docs/}. These changes have created significant challenges for researchers relying on these data sources to study social and cultural phenomena. 

\begin{figure*} [!h]
    \centering
    \includegraphics[width=1\linewidth]{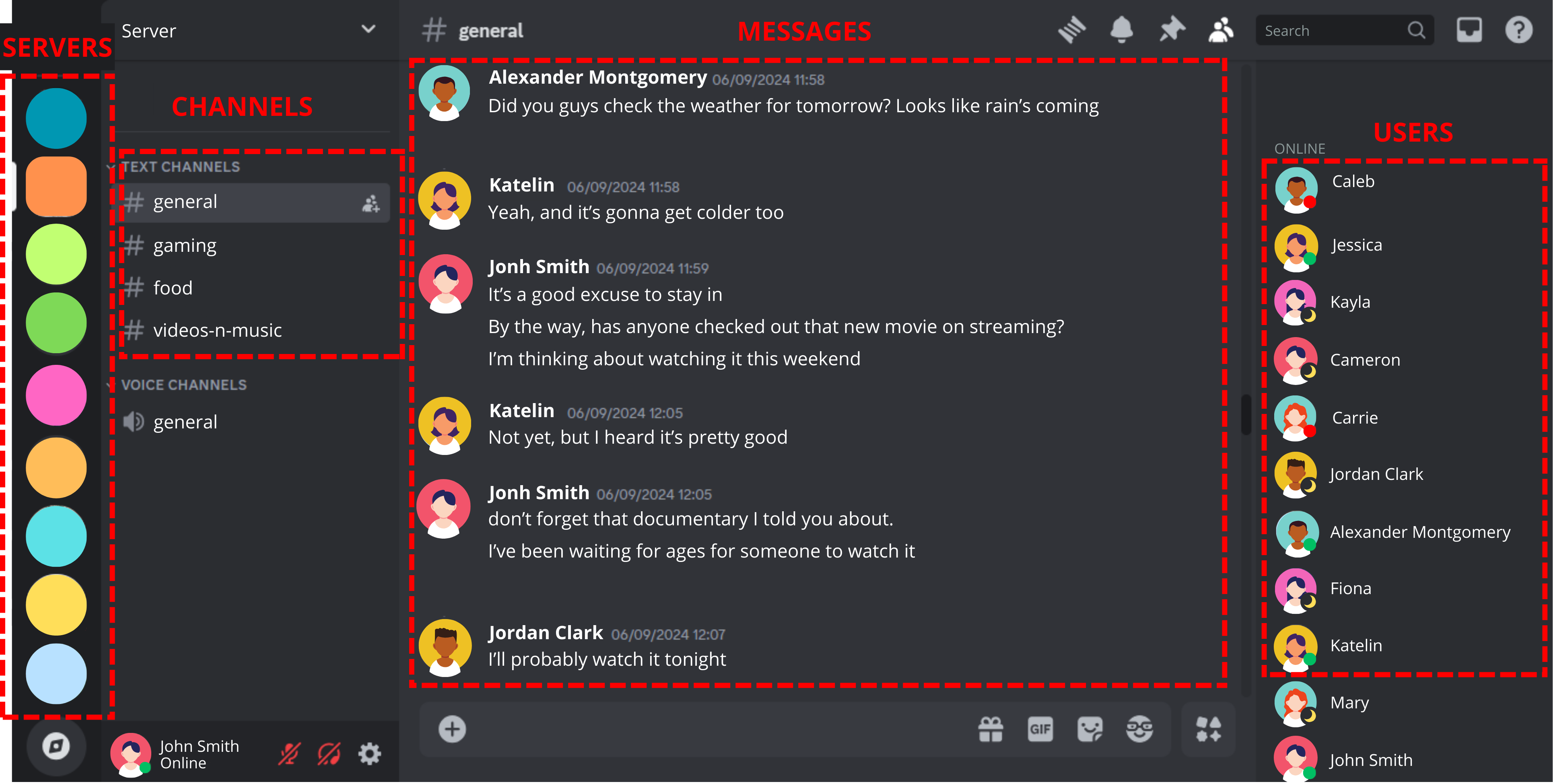}
    \caption{Discord interface, with the list of servers on the left, the distinct channels within the selected server, the central panel displaying messages from the active channel, and the list of connected users on the right.}
    \label{fig:disc}
\end{figure*} 

In contrast to these restrictive trends, data from Discord is available through its API. APIs provide a secure and standardized method for accessing data, ensuring compliance with user privacy guidelines and protection against unauthorized data collection practices. Discord's API allows researchers to collect data from public servers in a structured and ethical manner, adhering to the platform’s policies. This enables the study of large-scale interactions without compromising user privacy or data integrity.

While its API provides researchers with valuable access to public data, Discord’s versatility as a platform further amplifies its research potential. Initially developed as a communication tool for gamers, Discord has transformed into a dynamic platform that brings together communities with a wide range of interests \cite{johnson2022embracing}. Alongside private communication, its public servers host thousands of members, facilitating open conversations. This structure makes it a valuable context for studying social dynamics and digital communities, particularly in public settings where data can be collected. Despite being an emerging platform with an enormous user base and highly active communities, there remains a significant gap in works and analyses regarding Discord, with far fewer studies compared to other social networks.

In this regard, this paper introduces the most extensive Discord dataset available to date, comprising 2,052,206,308 messages from 4,735,057 unique users across 3,167 servers -- approximately 10\% of the servers listed in Discord's Discovery tab, a feature designed to highlight public servers that users can join. The dataset spans messages from 2015 to the end of 2024, capturing a wide variety of user interactions and community dynamics. By focusing on public servers, this dataset provides a robust and diverse foundation for exploring Discord as a unique social platform, offering insights into its distinctive structure and vibrant community-driven interactions. All data collection adhered strictly to Discord's API guidelines, and anonymization techniques were applied to ensure compliance with privacy standards.

This dataset marks a significant contribution to the study of digital communities, particularly in unmoderated and decentralized social platforms. Its scale and diversity enable comprehensive investigations into key topics such as governance models, moderation strategies, and the dissemination of information within dynamic online environments. Furthermore, it facilitates meaningful comparisons between Discord and traditional social networks, highlighting the unique interaction patterns fostered by community-driven moderation. The availability of this dataset opens new avenues for both theoretical and practical advancements, including the development of tools to analyze moderation policies, detect harmful behaviors, and model large-scale social interactions. By presenting this resource, we aim to cover the existing gap in research and studies on Discord, providing a valuable foundation for advancing the understanding of this increasingly relevant platform within computational social science.

%By providing this dataset, we contribute to the understanding of emerging social dynamics in decentralized platforms, enabling comparisons with traditional social networks and fostering research into community governance, moderation practices, and information dissemination. This study also highlights the role of secure and accessible APIs as a critical tool for computational social science research, especially as access to data on traditional platforms becomes increasingly limited. \yan{acho q da pra refazer esse final pra falar mais da relevancia do dataset e possíveis usos}

\section{What is Discord?}

Discord is a communication platform launched in 2015 by Discord Inc., designed to enable real-time interaction through text, voice, and video. Initially popular within gaming communities, Discord has expanded to accommodate diverse user groups, including educators, hobbyists, and professional teams\footnote{https://discord.com/safety/360044149331-what-is-discord}. Similar to platforms like Telegram and WhatsApp, Discord allows users to create and join customizable group spaces for synchronous and asynchronous communication, which can be either public or private.

Discord is organized into ``servers", virtual spaces designed for communities to connect, share content, and engage in conversations. Each server is divided into ``channels," which are dedicated spaces for specific types of conversations or activities. Text channels are used for written discussions, enabling members to exchange messages, share links, images, and other media. Voice channels, on the other hand, facilitate real-time audio conversations, often used for meetings, gaming sessions, or casual chats, and can also support video and screen sharing. All of these dynamics can be seen in Figure \ref{fig:disc}. Additionally, servers can be customized with ``roles" which are permission-based assignments that define what actions members can perform within the server. Roles can grant or restrict access to specific channels, allow users to manage messages, ban or mute members, or even adjust server settings. Combined with bots to automate tasks and server-specific rules to guide member behavior, this flexible architecture supports a wide range of communities, from small friend groups to large public forums.

A distinguishing feature of Discord is its user-driven moderation. While traditional social networks such as Twitter, Facebook, or YouTube have historically relied on centralized moderation managed by the platform itself \cite{wilson2020hate}, there is a very recent trend toward decentralizing moderation, exemplified by initiatives like community-based fact-checking \cite{balasubramanian2024publicdatasettrackingsocial}. Discord has always explicitly delegated the responsibility of rule-setting, behavior management, and access control to server administrators and moderators. This long-standing approach has fostered a well-established environment of non-centralized moderation, offering valuable insights into the dynamics and challenges of this model as other platforms begin to adopt similar strategies. For instance, as shown by \cite{moderation-challenges}, this flexibility can also be exploited to facilitate the presence of extremist and hateful groups and networks.

Bots are another key element of Discord’s ecosystem, setting it apart from other social networks. Unlike bots on other platforms, which are often limited to automated content posting or analytics, Discord bots are highly customizable and play an interactive role \cite{moderation-discord-bots, bots-discord}. They can automate moderation tasks, provide community engagement features (e.g., games, polls, or reminders), and integrate external services like music streaming, analytics tools, and generative AI services like Midjourney and ChatGPT. Discord actively encourages the creation and integration of bots by providing developers with extensive API documentation, support, and a bot-friendly environment, fostering innovation and customization within the platform \footnote{https://discord.com/developers/docs/intro}. Their extensive use has become central to how users interact within servers and how these communities are structured and managed.

To facilitate the discovery of public servers, Discord introduced the \textit{Server Discovery} feature. This feature allows users to search, explore and join public servers that align with their interests\footnote{https://support.discord.com/hc/en-us/articles/360023968311-Server-Discovery}. Servers listed in the Discovery tab must adhere to specific guidelines, including maintaining a welcoming environment, avoiding graphic or sexual content, and having an accurate server name and description\footnote{https://support.discord.com/hc/en-us/articles/4409308485271-Discovery-Guidelines}. These servers are the focus of our research, as they offer a curated look at communities
built around a wide variety of topics, each with at least 1000 members.
%with public servers being discoverable through Discord’s Discovery feature, provided they meet specific guidelines regarding content and moderation. 
% This work focuses exclusively on public Discord servers available in the Discovery feature
 %, from gaming and technology to education and entertainment.

%\yan{add um ultimo paragrafo?}

\section{Related Work}

% \pr{talvez o related work esteja grande, se nao tiver espaço no final, da pra tirar daqui}

% \subsection{Datasets in Social Media Research}

The availability of comprehensive datasets has underpinned advancements in computational social science, allowing researchers to explore user behavior, discourse, and network structures. Initially, research was strongly based on Twitter and Facebook data, given its initial openness to data access \cite{dooms2013movietweetings,lewis2008tastes}. After these platforms restricted access to most of their data, other platforms such as Reddit and WhatsApp became the focus of online social research. 

The Pushshift Reddit Dataset \cite{baumgartner2020pushshiftredditdataset}, for example, provides a detailed archive of Reddit posts and comments, facilitating studies of public discourse and online community dynamics. Similarly, GDELT \cite{Leetaru13gdelt:global} tracks global news events, offering tools for studying media trends, geopolitical developments, and event-based analyses. Also, WhatsApp datasets \cite{Seufert2023} have facilitated analyses of private communication in mobile-based social environments, contributing to the understanding of interpersonal dynamics.

Efforts have also focused on datasets from niche platforms, which provide unique insights into specific communities and communication structures. For instance, datasets from Twitter have been pivotal during specific events, such as the COVID-19 pandemic, enabling studies on information dissemination, public sentiment, and the spread of misinformation \cite{https://doi.org/10.5281/zenodo.7834392}. Similarly, recent datasets from Telegram public channels \cite{https://doi.org/10.5281/zenodo.7640712} and Koo posts, comments and profiles \cite{koo_Mekacher_Falkenberg_Baronchelli_2024} offer valuable perspectives on large-scale group interactions and content sharing. 

Social and news media datasets have further highlighted the potential for leveraging large-scale textual and visual data for societal impact studies. For instance, drought-related datasets \cite{Shang_Chen_Vora_Zhang_Cai_Wang_2024} integrate social media and news media to analyse environmental and socioeconomic effects. Similarly, datasets collected during the Russia-Ukraine conflict demonstrate the role of social media in information warfare and propaganda dissemination \cite{Ai_Gupta_Oak_Hui_Liu_Hirschberg_2024}. These datasets illustrate how social media data can be harnessed to address global and societal challenges.

Despite the breadth of available datasets, platforms like Discord remain underexplored. Discord’s focus on semi-private, community-oriented communication presents a distinct research opportunity. Unlike Twitter or Reddit, where public or semi-public interactions dominate, Discord facilitates structured, real-time conversations within server-based communities.

While prior works, such as \citet{Singh_Ghafouri_Such_Suarez-Tangil_2024} and \citet{10.1007/978-3-031-42171-6_5}, have introduced datasets of Discord messages, our work significantly expands the scope. We present the first general-purpose dataset, offering a far broader and more representative foundation for research. Building on a similar methodology, our dataset provides a more general and holistic perspective, addressing a critical gap in the research landscape. It is designed to enable studies of digital communication, community governance, moderation practices, cross-platform comparisons, and much more, thus unlocking new avenues for social media research.

\section{Dataset Construction}

\begin{figure*} [!ht]
    \centering
    \includegraphics[width=1\linewidth]{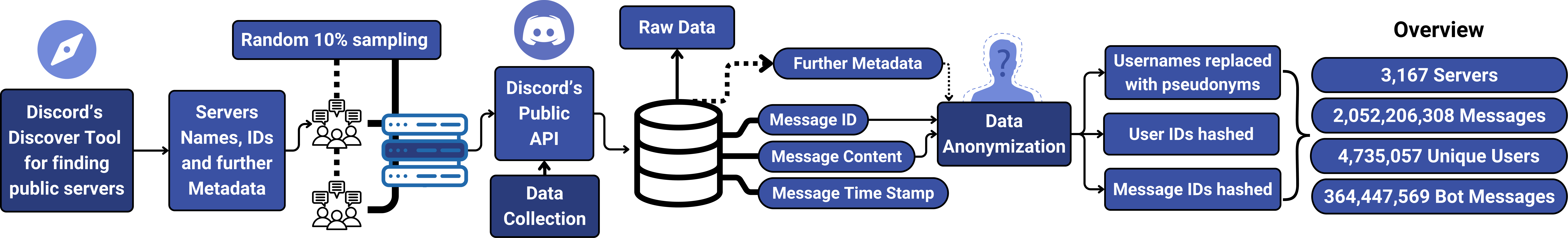}
    \caption{Diagram showcasing each step of the data collection process.}
    \label{fig:diagrama}
\end{figure*}

The dataset was built in three main phases: data collection, anonymization and organization, as detailed in this section.

\subsection{Data Collection}
\label{data-collection}
To access data from Discord servers, we first used Discord's Discovery feature, which allows users to browse public servers, and even see the public messages, without joining them. By querying the URL \textit{https://discord.com/servers?query=\{query\}}, where \textit{\{query\}} represents all possible combinations of letters and numbers of length up to four (which was the smallest length sufficient to retrieve all listed servers)
we systematically gathered server IDs, names, descriptions, and further metadata (e.g., server categories, keywords, and member count) for all servers available in Discovery. All of the metadata collected from each server is described in detail in the Appendix. Using this approach, we identified a total of 31,673 servers available in Discovery as of November 17, 2024.

% To access data from Discord servers, we first utilized the platform's Discovery feature, which allows users to explore public servers without needing to join them, to collect server IDs. Using this feature, we repeatedly queried the URL \textit{https://discord.com/servers?query=\textless query\textgreater}, where \textit{\textless query\textgreater} represents all possible combinations of letters and numbers of length four. This process allowed us to gather server IDs, names, descriptions, and additional metadata(e.g., server categories, keywords, and member count) for all public servers with at least 1,000 members. The specific metadata attributes collected for each server are detailed in Appendix \ref{tal}. Using this approach, we identified and recorded a total of 31,673 servers available in Discovery as of November 17, 2024.

Given the scale of the collected data, we implemented a random sampling strategy, selecting 10\% of all servers listed in Discord's Discovery tab. This sample, consisting of 3,167 servers, was chosen to balance feasibility with representativeness. The decision to limit the scope was motivated by computational and storage constraints, as well as adherence to Discord's API guidelines. Despite these limitations, we believe the sample size is sufficient to provide a meaningful characterization of the communities in Discord.

Using the server IDs from the sampled servers, we initiated the data collection process via the Discord API. First, we retrieved the publicly accessible channels for each server using the endpoint \textit{https://discord.com/api/v9/guilds/ \{server\_id\}/channels}. For each identified channel, we accessed all associated text messages by querying \textit{https://discord.com/api/v9/channels/\{channel\_id\}/messages}. The collection spanned from December 17, 2024, to January 2, 2025, ensuring that the dataset captured recent server activity during this period.

The complete workflow for data collection is illustrated in Figure \ref{fig:diagrama}. Further details on the attributes of the collected data are provided in the Data Instance Description section and in the Appendix.

\subsection{Anonymization Process}

Working with sensitive data that involves individuals requires addressing several ethical concerns. Even when data originate exclusively from public servers and messages, ensuring user anonymity remains a critical priority. This process aligns with principles such as GDPR and other privacy regulations, aiming to safeguard individual's identities while maintaining the utility of the dataset.

To address these concerns, our solution employs anonymization techniques to obscure sensitive information while preserving the dataset's analytical utility. Usernames are replaced with consistent pseudonyms generated by the \textit{mimesis library}, ensuring that identifiers remain unique and contextually meaningful across records. Similarly, user IDs and message IDs are hashed using the SHA-256 algorithm and truncated to 12 characters. This deterministic hashing approach maintains linkage between related records while effectively masking the original identifiers. The \textit{global\_name} field, deemed unnecessary for analysis, is entirely removed. Additionally, user IDs embedded within the content field are identified via regular expressions and replaced with their corresponding hash values. 

While the methods used significantly reduce the risk of reidentification, no anonymization process can guarantee absolute anonymity, particularly in datasets with rich contextual information. By sharing the detailed methodology, we aim to encourage transparency, enable validation, and promote ethical data usage. This anonymization process strikes a balance between preserving user privacy and maintaining the analytical value of the data set, ensuring that it can be safely used in future research.

\subsection{Dataset Organization}

The dataset was organized into individual \textbf{JSON files}, where each file corresponds to a specific server. The files are named after the servers ID's, ensuring reliability and consistency, and avoiding ambiguity or possible errors due to special characters.

Within each file, messages are stored in a format that prioritizes both channel organization and chronological order. Messages from the same channel are grouped, and within each channel, ordered from the most recent to the oldest.

This organization was designed to facilitate the analysis of temporal and channel-specific communication dynamics across servers. The details of the content within each message are described next.

\subsection{Data Description}

We opted to collect every single \textit{``Field"} made available by Discord's API. Each message is treated as an object with several attributes on our JSON files, and we also added a lone file to the dataset named ``servers\_metadata", which aggregates server related information from every server we had a message collected from. 

In this section, we introduce the structure and some details on those data instances to promote better interpretability.

\subsubsection{Message Object}

The main attributes of the message object are those commonly available in most chat-related datasets, such as: \textit{Message Content, Author (User ID, Username), Channel (Name, ID), Timestamp}.

However, many more sophisticated fields give access to other insightful information from each message, revealing aspects such as user behaviour or interactions. Examples include: \textit{``isBot"} indicates whether a message was sent from a bot or a personal account, \textit{``reactions?"} provides an array of reactions, a feature from the platform that allows users to interact with a particular message, \textit{``pinned"} reveals if a message was pinned or not, and \textit{``tts"} whether the message was transcribed through   `` text to speech" technology. All fields are listed in Table \ref{tab:discord_message_fields} in the Appendix. 

Messages are also classified according to the categories \textit{Message Types} and \textit{Flags}, including \textit{``DEFAULT"}, \textit{``REPLY"}, \textit{``USER\_JOIN"}, \textit{``IS\_CROSSPOST"} and many more.

\subsubsection{Server Metadata File}

Each server is also represented as an object with several attributes, capturing a wide range of metadata provided by Discord's Discovery Feature.

The main attributes include those essential for server identification and activity analysis, such as \textit{Server ID, Name, Description, Icon URL, Member Count}. Additional attributes allow for a deeper exploration of server characteristics and functionalities. For example, \textit{``slug"} provides a unique, URL-friendly identifier for the server, \textit{``preferred\_locale"} specifies the server's primary language, and \textit{``features"} lists special functionalities enabled for the server. Furthermore, fields like \textit{``reasons\_to\_join"} and \textit{``social\_links"} highlight promotional aspects and external connections of the server. 

Servers also include activity-related attributes such as \textit{``approximate\_presence\_count"} (the estimated number of currently active users) and \textit{``premium\_subscription\_count"} (indicating the number of premium subscriptions within the server). 
A complete list of fields and their descriptions can be found in Table \ref{tab:discord_server_fields} in the Appendix.

\subsubsection{Documentation} 

We highly encourage anyone interested in working with the dataset to go further and access Discord's official Documentation\footnote{https://discord.com/developers/docs/resources/message}, which provides more detailed information on the fields and attributes, ensuring a clearer understanding of the data structure and enabling more effective and thoughtful usage of the dataset for various purposes.

\subsection{Ethical Concerns}

Throughout every step of our data collection process, we prioritized adherence to ethical standards. Precautions were taken to collect data responsibly. All data was sourced from groups that are explicitly considered public according to Discord's terms of use, which every user agrees to upon signing up. The data was anonymized, and the methodology was detailed to promote reproducibility and transparency. Additionally, the data collection methods align with practices described in the current literature \cite{webmedia, Singh_Ghafouri_Such_Suarez-Tangil_2024, hateful_messages_adolescents}.

%All collected groups had at least 1,000 members and were accessible via Discord's ``Discovery" feature, which allows any user to view server content without joining. This ensures that the servers are unequivocally public. Furthermore, no interactions were made during the data collection process. We did not engage with any server or their members. Instead, Discord's API and server IDs were used to collect the data passively and non-intrusively.

Finally, the dataset released in this study adheres to the FAIR guiding principles for scientific data management and stewardship:

\textbf{Findable:} A unique and persistent Digital Object Identifier (DOI) is assigned to our dataset, ensuring it can be easily located and cited.

\textbf{Accessible:} The dataset is openly available for download, subject to the appropriate terms of use.

\textbf{Interoperable:} The dataset is released in JSON format, ensuring compatibility with a wide range of programming languages and operating systems.

\textbf{Reusable:} A detailed schema is provided in the Data Description section, clearly defining each field's purpose and facilitating reuse across diverse applications.

\section{Data Availability}
The dataset presented in this study has been made publicly available and can be accessed via DOI: 10.5281/zenodo.14658505\footnote{https://zenodo.org/records/14658505}. The data is provided in a compressed format, which can be decompressed for analysis. Detailed instructions for accessing and utilizing the dataset are provided in this article and the platform.

\section{Dataset Characterization}

\begin{figure*} [t]
    \centering
    \includegraphics[width=1\linewidth]{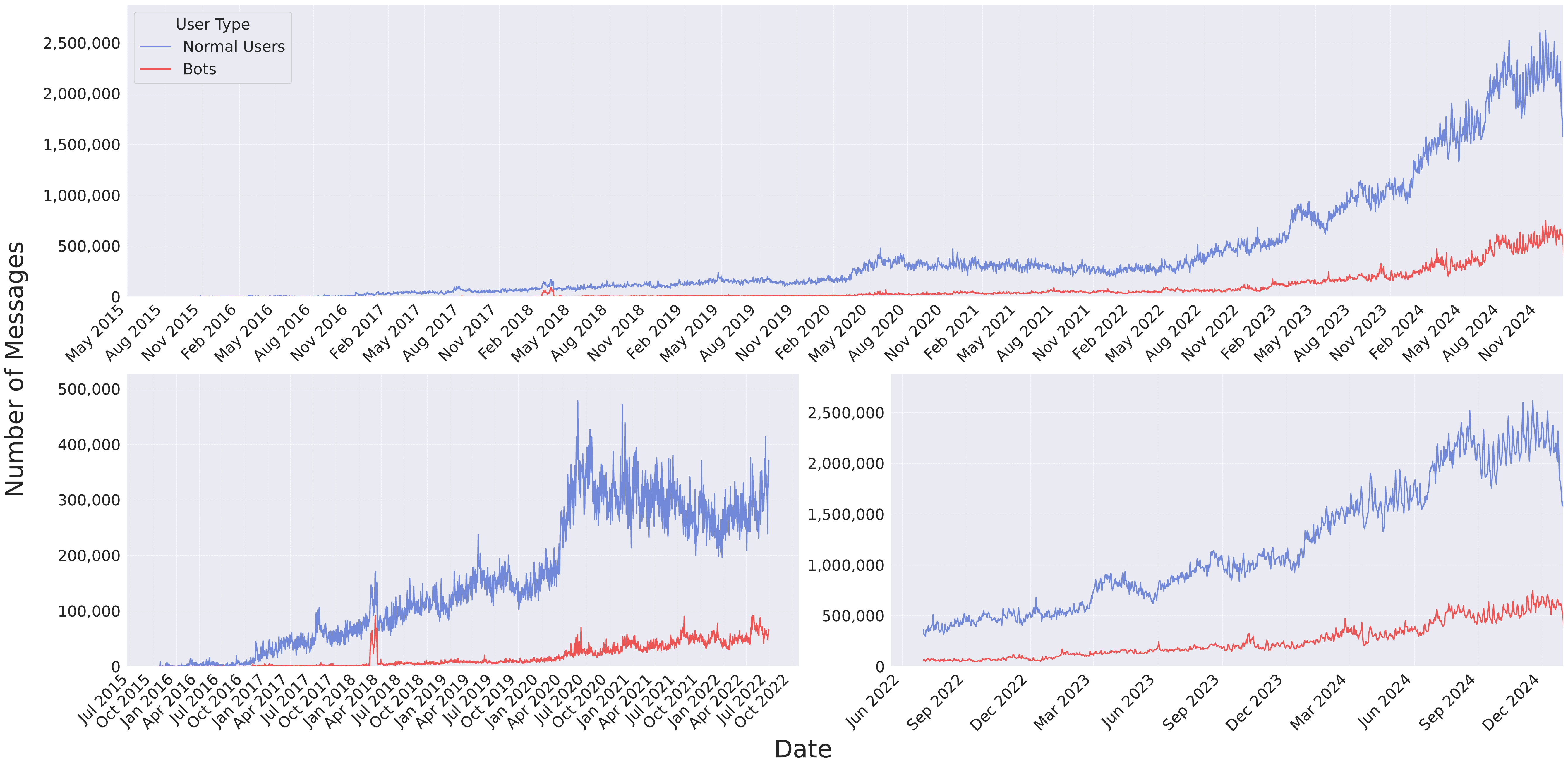}
    \caption{Evolution of the daily number of messages sent over time. The top panel presents the complete time series, distinguishing between messages sent by regular users (in blue) and bots (in red). The bottom-left panel focuses on the initial period, covering up to October 2022, while the bottom-right panel highlights the most recent period, from June 2022 to January 2024. Notice the graph scales are different.}
    \label{fig:messages_per_time}
\end{figure*}

% \gi{sugestao: fazer dos proximos paragrafos o inicio da secao de caracterizacao.}

The \textbf{3,167 collected servers} yielded a total of \textbf{2,052,206,308 unique messages} sent by \textbf{4,735,057 distinct users}. From the total number of messages,   \textbf{364,447,569 (17\%) originated from bots}. The data spans from Discord's launch, on \textbf{May 13, 2015}, to \textbf{December 17, 2024}, when the data collection process began.

Figure \ref{fig:messages_per_time} shows the number of messages over time. 
Note that Discord's retroactive approach to chat allows the dataset to encompass a timeframe broader than the actual data collection period.
Upon joining a server, users gain access to all non-deleted historical content within public channels, and the same is valid for data retrieval using their API.
Notably, 2024 stands out as the most active year across all servers, reflecting a growing network that offers ample opportunities for further exploration.

Although it is interesting to be able to access older data, it is important to note that the servers listed in the \textit{Discovery} tab most likely represent those active at the time of data collection. Servers featured in this tab typically emphasize ongoing engagement and relevance. Additionally, while Discord provides access to historical content, some servers may periodically delete older messages, which could introduce variability in the completeness of the temporal data available. These factors may influence the representativeness of older content and should be considered when analysing the dataset.

Next, we present simple but relevant analyses of the dataset. The first aspect we examined is \textbf{Bots.} As previously mentioned, Discord bots play a crucial role in enhancing user experience by automating tasks, facilitating engagement, and offering specialized functionalities. They serve as virtual assistants that can perform various tasks, such as moderation, entertainment, and utility services. Table \ref{tab:top10bots} illustrates the diverse applications of bots, highlighting their role in generating messages, being mentioned by users, and eliciting reactions, which are key indicators of their impact and utility in Discord.

Bots like \textit{MEE6} and \textit{Dyno} are widely recognized as powerful moderators, enabling server administrators to enforce rules, assign roles, and monitor activity effectively. These bots are essential for maintaining order in larger communities where manual moderation would be impractical. Similarly, \textit{Arcane Premium} and \textit{Loritta} excel in general server management, providing features such as levelling systems, customizable commands, and automated event handling to enhance user engagement and server functionality. On the other hand, entertainment-focused bots like \textit{Pokétwo}, \textit{Mudae}, \textit{Karuta} and \textit{OwO} captivate users with gaming and collectable experiences, encouraging interaction through game mechanics such as Pokémon battles, waifu collections, trading cards and interacting with fictional animals. These bots create an environment where users actively participate, forming micro-communities within the larger server ecosystem. Finally, \textit{Lucky VR}, despite being the bot with the highest volume of messages, lacks extensive documentation online. Its primary focus is on gambling in virtual reality, particularly in facilitating poker games.

\begin{figure}[t]
    \centering
    \includegraphics[width=1\linewidth]{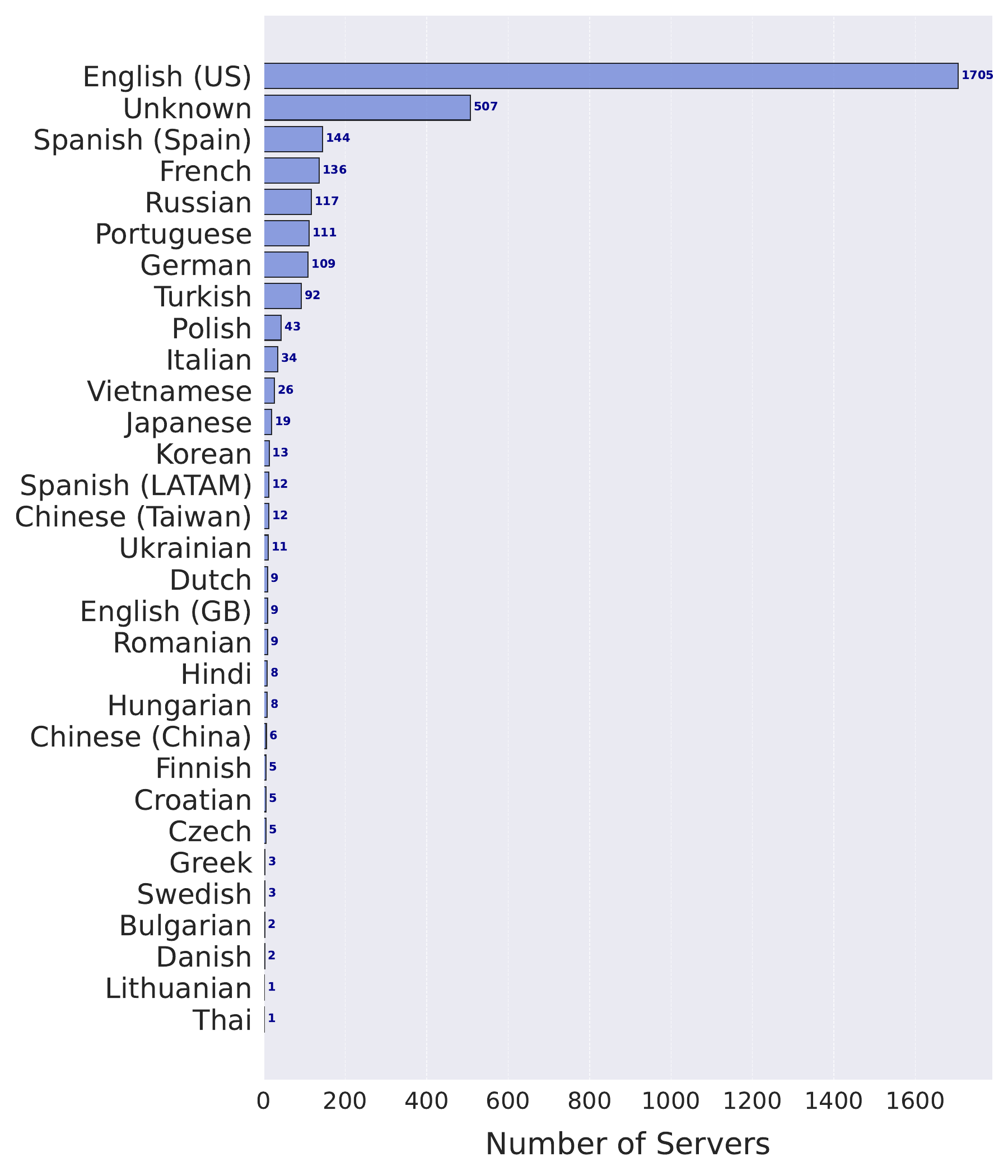}
    \caption{Bar plot of the number of servers by language, with the majority of servers being in English.}
    \label{fig:hist-linguas}
\end{figure}

\textbf{Languages.} To identify the most frequent languages in our dataset, we analysed the values present in the \textit{preferred locale} field within the server metadata. Figure \ref{fig:hist-linguas} shows English (US) as the primary language for most servers, with 1,705 servers. Disregarding the ``unknown" field, the second most frequent language is Spanish (Spain), with 144 servers, followed by French, with 136 servers. Also note there is a notable linguistic diversity, including Portuguese, Russian, and German.

% For instance, bots like MEE6 and Dyno are widely used for community management, handling tasks like assigning roles, enforcing rules, and moderating content. On the other hand, entertainment bots such as Mudae, Karuta, and Pokétwo engage users through games, collectibles, and interactive features, exhibiting higher levels of participation and interaction. Additionally, niche bots like AniLibria.TV provide specialized services, such as streaming updates or content delivery for targeted audiences. The bots listed in Table \ref{tab:top10bots} illustrate the diverse applications of Discord bots, highlighting their role in generating messages, being mentioned by users, and eliciting reactions, which are key indicators of their impact and utility in various online communities.

\begin{table*}[t]
\caption{Top 10 Bots by Messages, Mentions, and Reactions.}
\centering
\label{tab:top10bots}
\begin{tabular}{ccc}
\toprule
\textbf{Messages} & \textbf{Mentions} & \textbf{Reactions} \\
\midrule
\hline
\begin{tabular}{lr}
Lucky VR & 3,684,600 \\
Mudae & 2,955,892 \\
AniLibria.TV & 2,844,343 \\
Pok\'etwo & 2,005,250 \\
RoM & 1,716,654 \\
OwO & 1,523,404 \\
Mimu & 1,430,137 \\
Wan Shi Tong & 1,297,316 \\
BTE France Minecraft & 1,077,553 \\
MEE6 & 904,432 \\
\end{tabular} &
\begin{tabular}{lr}
Pok\'etwo & 304,657 \\
Dyno & 290,387 \\
AniLibria.TV & 290,286 \\
ProBot  & 278,653 \\
Loritta & 251,208 \\
Arcane Premium & 224,681 \\
Mimu & 206,157 \\
MEE6 & 198,534 \\
Dank Memer & 178,813 \\
Liquid Esports & 134,816 \\
\end{tabular} &
\begin{tabular}{lr}
Karuta & 984,693 \\
Mudae & 281,929 \\
MEE6 & 237,370 \\
XenosPD.dk & 217,634 \\
Bongo & 216,779 \\
OwO & 193,493 \\
YAGPDB.xyz & 137,052 \\
Dyno & 131,881 \\
Suggester & 88,981 \\
Emps-World & 85,889 \\
\end{tabular} \\
\bottomrule
\hline
\end{tabular}
\end{table*}

\textbf{Servers.} Discord, as a platform for community interactions, supports a wide range of themes that reflect different user interests. To analyze the thematic distribution of servers, we examined server metadata, focusing on keywords that describe their content. These keywords also improve the platform's search functionality through the Discovery feature.

Table \ref{tab:top_40_keywords} presents the 40 most frequent keywords appearing in the description field of the servers collected. Despite Discord's evolution into a versatile platform, gaming-related content remains significant, with \textit{gaming} appearing in over 15\% of descriptions. Related keywords like \textit{minecraft}, \textit{roblox}, and \textit{twitch} reinforce its gaming's central role. Other prominent themes include \textit{anime} (5.68\%), \textit{roleplay} (6.44\%), and \textit{fivem} (4.39\%), reflecting niche communities and interactive multiplayer experiences. 

Emerging interests like \textit{esports} (4.04\%) and \textit{social} (4.14\%) highlight the importance of competitive gaming and socialization. Creative and recreational themes, such as \textit{art} and \textit{music}, and educational discussions, such as \textit{programming}, demonstrate the platform's versatility beyond gaming.

\begin{table}[t]
\caption{Top 40 most frequent keywords in Discord servers description and their coverage.}
\centering
\begin{tabular}{lr|lr}
%\hline
%\multicolumn{4}{c}{\textbf{Keywords}} \\
\hline
\textbf{Keyword} & \textbf{\%} & \textbf{Keyword} & \textbf{\%} \\
\hline
gaming & 15.28 & giveaways & 2.31 \\
youtube & 15.00 & art & 2.02 \\
minecraft & 12.28 & rp & 2.02 \\
roblox & 10.83 & tiktok & 1.89 \\
twitch & 8.68 & manga & 1.83 \\
community & 8.65 & chill & 1.74 \\
roleplay & 6.44 & leagueoflegends & 1.71 \\
anime & 5.68 & rpg & 1.71 \\
fivem & 4.39 & pvp & 1.58 \\
social & 4.14 & music & 1.48 \\
esports & 4.04 & server & 1.39 \\
valorant & 3.79 & chatting & 1.36 \\
memes & 2.97 & survival & 1.36 \\
fortnite & 2.78 & gta5 & 1.33 \\
streamer & 2.59 & modding & 1.20 \\
events & 2.53 & mobile & 1.20 \\
game & 2.49 & gtav & 1.20 \\
fun & 2.46 & programming & 1.17 \\
gta & 2.43 & pc & 1.17 \\
games & 2.43 & csgo & 1.14 \\
\hline
\label{tab:top_40_keywords}
\end{tabular}
\end{table}

However, note that these keywords are only related to the server's descriptions. The context of messages goes beyond these subjects, with messages concerning topics that include mental health, political debates and web dating, for example. \looseness=-1

\section{Potential Applications}

Our dataset represents the largest publicly available collection of textual data from Discord servers, offering an unprecedented resource for studying online interactions. The dataset spans a wide range of languages, cultures, and topics, making it uniquely suited to support a diverse cross-section of research efforts and laying a robust foundation for examining the complexities of online communities. Its breadth and depth enable researchers to explore critical areas such as discourse analysis, community governance, and political debate, making it an invaluable asset for interdisciplinary studies.\looseness=-1

\textbf{Online Community Governance.} Discord's user-driven moderation model stands apart from those used in platforms like Facebook and YouTube, which rely heavily on centralized moderation enforced by platform administrators or automated systems \cite{doi:10.1177/1461444818821316}. This decentralized approach enables server owners and moderators to define and enforce community-specific rules, offering an opportunity to explore how self-governance influences online interactions, social dynamics, and conflict resolution \cite{moderationbook}. Researchers can examine the efficacy of decentralized moderation in fostering inclusive and safe environments, the strategies employed by communities to address toxic behavior, and the broader impact of granting users autonomy in governing digital spaces \cite{succesfulcommunitybook}.

\textbf{Discourse analysis.} The dataset also provides a valuable resource for advancing research across multiple scientific domains, particularly in Natural Language Processing (NLP) and Machine Learning (ML). It enables the development and evaluation of models for tasks such as sentiment analysis, intent recognition, topic modeling, and toxic or abusive language detection. Moreover, the temporal structure of the data supports studies on conversational dynamics, social network analysis, and community behavior in digital environments. This dataset can also facilitate the creation of domain-specific chatbots, recommendation systems, and tools for automated moderation, fostering innovations that bridge computational techniques with the study of human communication and online interaction.

\textbf{Political Debate.} The political debate research community faces substantial challenges in understanding how both mainstream and decentralized platforms influence political opinions and behaviors. Social networks have become increasingly critical in shaping political landscapes, not only by amplifying voices and ideologies but also by contributing to the rapid dissemination of fake news and misinformation \cite{Aimeur2023}. While much of the existing research focuses on platforms like Facebook and Twitter \cite{10.1145/3578503.3583597, doi:10.1126/science.adk3451}, Discord’s semi-private and community-driven architecture offers a distinct and underexplored environment for political conversations. 
%This dynamic has raised significant concerns about the potential for unregulated spaces to facilitate polarization, radicalization, and the erosion of trust in democratic processes \cite{moderation-challenges}. At the same time, Discord's user-moderated model provides a valuable opportunity to study these phenomena in a decentralized context, contrasting with the centralized moderation models of mainstream platforms. -> Achei que estava redundante com outra parte do texto
Our dataset enables researchers to explore the impact of digital platforms on political discourse, the propagation of misinformation, and the development of effective moderation and regulation strategies tailored to such environments.

\textbf{Mental health.} The relationship between social media usage and the prevalence of mental health issues, including self-harm and suicidal behavior, has been extensively documented, particularly among younger demographics \cite{elia2020, pater2016characterizations}. Discord, as a platform with diverse communities and user-generated content, represents a critical environment for understanding these phenomena. Recent studies in Brazil have already looked at mental health in the context of Discord, underscoring the importance of examining how the platform’s semi-private and community-oriented spaces may influence mental health outcomes \cite{webmedia}. Our multilingual dataset significantly expands the scope for such research by enabling cross-cultural and cross-linguistic analyses of mental health trends and discourse on Discord, providing a valuable foundation for identifying patterns of at-risk behavior and explore critical questions such as the prevalence of harmful behaviors or supportive interactions.

\section{Discussion and Conclusion}

This study introduces the Discord Unveiled Dataset, a comprehensive and ethically curated resource encompassing over 3,000 public servers and 2 billion messages exchanged on Discord. By leveraging Discord's API and implementing anonymization protocols, we ensured compliance with privacy standards while maintaining the analytical utility of the dataset. Spanning nearly a decade of public interactions, the dataset provides a robust foundation for investigating decentralized moderation, online community dynamics, and patterns of digital communication.

The dataset highlights the distinctive features of Discord as a platform, particularly its user-driven moderation model and the extensive use of bots to enhance functionality. These characteristics differentiate Discord from more centralized platforms like Facebook and Twitter, offering a novel perspective on how decentralized platforms shape online interactions. Additionally, the dataset's multilingual and multifaceted nature opens avenues for research on underexplored languages and cultures in digital spaces, addressing gaps in computational social science. 

However, the dataset still has limitations, most notably the absence of detailed user-level data, such as demographic information and activity patterns, which could enrich analyses of individual behavior and group dynamics. This limitation reflects the constraints of Discord's API and underscores the need for complementary data sources to achieve a more granular understanding. Additionally, it is essential to address potential risks and ethical concerns. The dataset could be misused to develop tools that amplify misinformation, harassment, or other harmful behaviors. To mitigate these risks, we adhered to strict ethical guidelines during dataset creation, employing robust anonymization protocols and privacy-compliance measures to protect user identities. %We encourage researchers to engage with ethical review processes, critically assess the societal implications of their work, and ensure responsible data use. 

Despite these challenges, we believe the Discord Unveiled Dataset will serve as a valuable resource for advancing computational and social science research, fostering a deeper understanding of the complexities of modern digital platforms. %By releasing this dataset, we aim to bridge the gap in research on Discord and offer a valuable resource that fosters a nuanced understanding of the complexities of contemporary digital platforms.

\section{Acknowledgments}

This work was partially funded by CNPq, CAPES, FAPEMIG, and IAIA - INCT on AI.

\bibliography{main}

\begin{thebibliography}{31}
\providecommand{\natexlab}[1]{#1}

\bibitem[{Abi-Jaoude, Naylor, and Pignatiello(2020)}]{elia2020}
Abi-Jaoude, E.; Naylor, K.~T.; and Pignatiello, A. 2020.
\newblock Smartphones, social media use and youth mental health.
\newblock \emph{Cmaj}, 192(6): E136--E141.

\bibitem[{Ai et~al.(2024)Ai, Gupta, Oak, Hui, Liu, and Hirschberg}]{Ai_Gupta_Oak_Hui_Liu_Hirschberg_2024}
Ai, L.; Gupta, S.; Oak, S.; Hui, Z.; Liu, Z.; and Hirschberg, J. 2024.
\newblock TweetIntent@Crisis: A Dataset Revealing Narratives of Both Sides in the Russia-Ukraine Crisis.
\newblock \emph{Proceedings of the International AAAI Conference on Web and Social Media}, 18(1): 1872--1887.

\bibitem[{Allen, Watts, and Rand(2024)}]{doi:10.1126/science.adk3451}
Allen, J.; Watts, D.~J.; and Rand, D.~G. 2024.
\newblock Quantifying the impact of misinformation and vaccine-skeptical content on Facebook.
\newblock \emph{Science}, 384(6699): eadk3451.

\bibitem[{Aïmeur, Amri, and Brassard(2023)}]{Aimeur2023}
Aïmeur, E.; Amri, S.; and Brassard, G. 2023.
\newblock Fake news, disinformation and misinformation in social media: a review.
\newblock \emph{Social Network Analysis and Mining}, 13(1): 30.

\bibitem[{Bakshy et~al.(2012)Bakshy, Rosenn, Marlow, and Adamic}]{facebook-1}
Bakshy, E.; Rosenn, I.; Marlow, C.; and Adamic, L. 2012.
\newblock The Role of Social Networks in Information Diffusion.
\newblock arXiv:1201.4145.

\bibitem[{Balasubramanian et~al.(2024)Balasubramanian, Zou, Narayana, You, Luceri, and Ferrara}]{balasubramanian2024publicdatasettrackingsocial}
Balasubramanian, A.; Zou, V.; Narayana, H.; You, C.; Luceri, L.; and Ferrara, E. 2024.
\newblock A Public Dataset Tracking Social Media Discourse about the 2024 U.S. Presidential Election on Twitter/X.
\newblock arXiv:2411.00376.

\bibitem[{Banda et~al.(2023)Banda, Tekumalla, Wang, Yu, Liu, Ding, Artemova, Tutubalina, and Chowell}]{https://doi.org/10.5281/zenodo.7834392}
Banda, J.~M.; Tekumalla, R.; Wang, G.; Yu, J.; Liu, T.; Ding, Y.; Artemova, K.; Tutubalina, E.; and Chowell, G. 2023.
\newblock A large-scale COVID-19 Twitter chatter dataset for open scientific research - an international collaboration.

\bibitem[{Baumgartner et~al.(2020)Baumgartner, Zannettou, Keegan, Squire, and Blackburn}]{baumgartner2020pushshiftredditdataset}
Baumgartner, J.; Zannettou, S.; Keegan, B.; Squire, M.; and Blackburn, J. 2020.
\newblock The Pushshift Reddit Dataset.
\newblock arXiv:2001.08435.

\bibitem[{Bento et~al.(2024)Bento, Buzelin, Aquino, Carvalho, Dutenhefner, Dayrell, Santana, Estanislau, Pappa, Miranda, Almeida, and Jr}]{webmedia}
Bento, P.; Buzelin, A.; Aquino, Y.; Carvalho, I.; Dutenhefner, P.; Dayrell, L.; Santana, C.; Estanislau, V.; Pappa, G.; Miranda, D.; Almeida, V.; and Jr, W.~M. 2024.
\newblock Impacto da Pandemia na Discussão sobre Saúde Mental: O Caso do Discord no Brasil.
\newblock In \emph{Proceedings of the 30th Brazilian Symposium on Multimedia and the Web}, 179--187. Porto Alegre, RS, Brasil: SBC.

\bibitem[{Dooms, De~Pessemier, and Martens(2013)}]{dooms2013movietweetings}
Dooms, S.; De~Pessemier, T.; and Martens, L. 2013.
\newblock Movietweetings: a movie rating dataset collected from twitter.
\newblock In \emph{Workshop on Crowdsourcing and human computation for recommender systems, CrowdRec at RecSys}, volume 2013, 43.

\bibitem[{Efstratiou et~al.(2023)Efstratiou, Blackburn, Caulfield, Stringhini, Zannettou, and De~Cristofaro}]{hate-speech-reddit-savvas}
Efstratiou, A.; Blackburn, J.; Caulfield, T.; Stringhini, G.; Zannettou, S.; and De~Cristofaro, E. 2023.
\newblock Non-polar Opposites: Analyzing the Relationship between Echo Chambers and Hostile Intergroup Interactions on Reddit.
\newblock \emph{Proceedings of the International AAAI Conference on Web and Social Media}, 17: 197--208.

\bibitem[{Fillies, Peikert, and Paschke(2023)}]{hateful_messages_adolescents}
Fillies, J.; Peikert, S.; and Paschke, A. 2023.
\newblock Hateful Messages: A Conversational Data Set of Hate Speech produced by Adolescents on Discord.

\bibitem[{Fillies, Peikert, and Paschke(2024)}]{10.1007/978-3-031-42171-6_5}
Fillies, J.; Peikert, S.; and Paschke, A. 2024.
\newblock Hateful Messages: A Conversational Data Set of Hate Speech Produced by Adolescents on Discord.
\newblock In Haber, P.; Lampoltshammer, T.~J.; and Mayr, M., eds., \emph{Data Science---Analytics and Applications}, 37--44. Cham: Springer Nature Switzerland.
\newblock ISBN 978-3-031-42171-6.

\bibitem[{Gillespie(2018)}]{moderationbook}
Gillespie, T. 2018.
\newblock \emph{Custodians of the Internet: Platforms, Content Moderation, and the Hidden Decisions That Shape Social Media}.
\newblock ISBN 9780300235029.

\bibitem[{{Hoevers}(2022)}]{bots-discord}
{Hoevers}, S. 2022.
\newblock Discourse on Discord : An analysis of relations between users.

\bibitem[{Johnson and Salter(2022)}]{johnson2022embracing}
Johnson, E.~K.; and Salter, A. 2022.
\newblock Embracing discord? The rhetorical consequences of gaming platforms as classrooms.
\newblock \emph{Computers and Composition}, 65: 102729.

\bibitem[{Kiene and Hill(2020)}]{moderation-discord-bots}
Kiene, C.; and Hill, B.~M. 2020.
\newblock Who Uses Bots? A Statistical Analysis of Bot Usage in Moderation Teams.
\newblock In \emph{Extended Abstracts of the 2020 CHI Conference on Human Factors in Computing Systems}, CHI EA '20, 1–8. New York, NY, USA: Association for Computing Machinery.
\newblock ISBN 9781450368193.

\bibitem[{Kraut and Resnick(2012)}]{succesfulcommunitybook}
Kraut, R.; and Resnick, P. 2012.
\newblock \emph{Building Successful Online Communities: Evidence-Based Social Design}.
\newblock ISBN 9780262298315.

\bibitem[{La~Morgia, Mei, and Mongardini(2023)}]{https://doi.org/10.5281/zenodo.7640712}
La~Morgia, M.; Mei, A.; and Mongardini, A.~M. 2023.
\newblock TGDataset.

\bibitem[{Leetaru and Schrodt(2013)}]{Leetaru13gdelt:global}
Leetaru, K.; and Schrodt, P.~A. 2013.
\newblock GDELT: Global data on events, location, and tone.
\newblock \emph{ISA Annual Convention}.

\bibitem[{Lewis et~al.(2008)Lewis, Kaufman, Gonzalez, Wimmer, and Christakis}]{lewis2008tastes}
Lewis, K.; Kaufman, J.; Gonzalez, M.; Wimmer, A.; and Christakis, N. 2008.
\newblock Tastes, ties, and time: A new social network dataset using Facebook. com.
\newblock \emph{Social networks}, 30(4): 330--342.

\bibitem[{Mekacher, Falkenberg, and Baronchelli(2024)}]{koo_Mekacher_Falkenberg_Baronchelli_2024}
Mekacher, A.; Falkenberg, M.; and Baronchelli, A. 2024.
\newblock The Koo Dataset: An Indian Microblogging Platform with Global Ambitions.
\newblock \emph{Proceedings of the International AAAI Conference on Web and Social Media}, 18(1): 1991--2002.

\bibitem[{Pater et~al.(2016)Pater, Kim, Mynatt, and Fiesler}]{pater2016characterizations}
Pater, J.; Kim, S.; Mynatt, E.; and Fiesler, C. 2016.
\newblock Characterizations of Online Harassment: Comparing Policies Across Social Media Platforms.
\newblock In \emph{Proceedings of the 19th International Conference on Supporting Group Work (GROUP '16)}, 369--374. New York, NY, USA: Association for Computing Machinery.

\bibitem[{Pierri et~al.(2023)Pierri, Luceri, Jindal, and Ferrara}]{10.1145/3578503.3583597}
Pierri, F.; Luceri, L.; Jindal, N.; and Ferrara, E. 2023.
\newblock Propaganda and Misinformation on Facebook and Twitter during the Russian Invasion of Ukraine.
\newblock In \emph{Proceedings of the 15th ACM Web Science Conference 2023}, WebSci '23, 65–74. New York, NY, USA: Association for Computing Machinery.
\newblock ISBN 9798400700897.

\bibitem[{Seering(2020)}]{moderation-challenges}
Seering, J. 2020.
\newblock Reconsidering Self-Moderation: the Role of Research in Supporting Community-Based Models for Online Content Moderation.
\newblock \emph{Proc. ACM Hum.-Comput. Interact.}, 4(CSCW2).

\bibitem[{Seering et~al.(2019)Seering, Wang, Yoon, and Kaufman}]{doi:10.1177/1461444818821316}
Seering, J.; Wang, T.; Yoon, J.; and Kaufman, G. 2019.
\newblock Moderator engagement and community development in the age of algorithms.
\newblock \emph{New Media \& Society}, 21(7): 1417--1443.

\bibitem[{Seufert et~al.(2023)Seufert, Poignée, Hoßfeld, and Seufert}]{Seufert2023}
Seufert, A.; Poignée, F.; Hoßfeld, T.; and Seufert, M. 2023.
\newblock {WhatsApp Data Set}.

\bibitem[{Shang et~al.(2024)Shang, Chen, Vora, Zhang, Cai, and Wang}]{Shang_Chen_Vora_Zhang_Cai_Wang_2024}
Shang, L.; Chen, B.; Vora, A.; Zhang, Y.; Cai, X.; and Wang, D. 2024.
\newblock SocialDrought: A Social and News Media Driven Dataset and Analytical Platform towards Understanding Societal Impact of Drought.
\newblock \emph{Proceedings of the International AAAI Conference on Web and Social Media}, 18(1): 2051--2062.

\bibitem[{Singh et~al.(2024)Singh, Ghafouri, Such, and Suarez-Tangil}]{Singh_Ghafouri_Such_Suarez-Tangil_2024}
Singh, A.~K.; Ghafouri, V.; Such, J.; and Suarez-Tangil, G. 2024.
\newblock Differences in the Toxic Language of Cross-Platform Communities.
\newblock \emph{Proceedings of the International AAAI Conference on Web and Social Media}, 18(1): 1463--1476.

\bibitem[{Wang et~al.(2023)Wang, Luceri, Pierri, and Ferrara}]{moderation-importance-qanon}
Wang, E.~L.; Luceri, L.; Pierri, F.; and Ferrara, E. 2023.
\newblock Identifying and Characterizing Behavioral Classes of Radicalization within the QAnon Conspiracy on Twitter.
\newblock arXiv:2209.09339.

\bibitem[{Wilson and Land(2020)}]{wilson2020hate}
Wilson, R.~A.; and Land, M.~K. 2020.
\newblock Hate speech on social media: Content moderation in context.
\newblock \emph{Conn. L. Rev.}, 52: 1029.

\end{thebibliography}

\subsection{Paper Checklist}

\begin{enumerate}

\item For most authors...
\begin{enumerate}
    \item  Would answering this research question advance science without violating social contracts, such as violating privacy norms, perpetuating unfair profiling, exacerbating the socio-economic divide, or implying disrespect to societies or cultures?
    \answerYes{Yes}
  \item Do your main claims in the abstract and introduction accurately reflect the paper's contributions and scope?
    \answerYes{Yes}
   \item Do you clarify how the proposed methodological approach is appropriate for the claims made? 
    \answerYes{Yes}
   \item Do you clarify what are possible artifacts in the data used, given population-specific distributions?
    \answerYes{Yes}
  \item Did you describe the limitations of your work?
    \answerYes{Yes}
  \item Did you discuss any potential negative societal impacts of your work?
    \answerYes{Yes}
      \item Did you discuss any potential misuse of your work?
    \answerYes{Yes}
    \item Did you describe steps taken to prevent or mitigate potential negative outcomes of the research, such as data and model documentation, data anonymization, responsible release, access control, and the reproducibility of findings?
    \answerYes{Yes}
  \item Have you read the ethics review guidelines and ensured that your paper conforms to them?
    \answerYes{Yes}
\end{enumerate}

\item Additionally, if your study involves hypotheses testing...
\begin{enumerate}
  \item Did you clearly state the assumptions underlying all theoretical results?
    \answerNA{NA}
  \item Have you provided justifications for all theoretical results?
    \answerNA{NA}
  \item Did you discuss competing hypotheses or theories that might challenge or complement your theoretical results?
    \answerNA{NA}
  \item Have you considered alternative mechanisms or explanations that might account for the same outcomes observed in your study?
    \answerNA{NA}
  \item Did you address potential biases or limitations in your theoretical framework?
    \answerNA{NA}
  \item Have you related your theoretical results to the existing literature in social science?
    \answerNA{NA}
  \item Did you discuss the implications of your theoretical results for policy, practice, or further research in the social science domain?
    \answerNA{NA}
\end{enumerate}

\item Additionally, if you are including theoretical proofs...
\begin{enumerate}
  \item Did you state the full set of assumptions of all theoretical results?
    \answerNA{NA}
	\item Did you include complete proofs of all theoretical results?
    \answerNA{NA}
\end{enumerate}

\item Additionally, if you ran machine learning experiments...
\begin{enumerate}
  \item Did you include the code, data, and instructions needed to reproduce the main experimental results (either in the supplemental material or as a URL)?
    \answerNA{NA}
  \item Did you specify all the training details (e.g., data splits, hyperparameters, how they were chosen)?
    \answerNA{NA}
     \item Did you report error bars (e.g., with respect to the random seed after running experiments multiple times)?
    \answerNA{NA}
	\item Did you include the total amount of compute and the type of resources used (e.g., type of GPUs, internal cluster, or cloud provider)?
    \answerNA{NA}
     \item Do you justify how the proposed evaluation is sufficient and appropriate to the claims made? 
    \answerNA{NA}
     \item Do you discuss what is ``the cost`` of misclassification and fault (in)tolerance?
    \answerNA{NA}
  
\end{enumerate}

\item Additionally, if you are using existing assets (e.g., code, data, models) or curating/releasing new assets, \textbf{without compromising anonymity}...
\begin{enumerate}
  \item If your work uses existing assets, did you cite the creators?
    \answerNA{NA}
  \item Did you mention the license of the assets?
    \answerNA{NA}
  \item Did you include any new assets in the supplemental material or as a URL?
    \answerNA{NA}
  \item Did you discuss whether and how consent was obtained from people whose data you're using/curating?
    \answerYes{Yes}
  \item Did you discuss whether the data you are using/curating contains personally identifiable information or offensive content?
    \answerYes{Yes}
\item If you are curating or releasing new datasets, did you discuss how you intend to make your datasets FAIR?
\answerYes{Yes}
\item If you are curating or releasing new datasets, did you create a Datasheet for the Dataset (see \citet{gebru2021datasheets})? 
\answerYes{Yes}
\end{enumerate}

\item Additionally, if you used crowdsourcing or conducted research with human subjects, \textbf{without compromising anonymity}...
\begin{enumerate}
  \item Did you include the full text of instructions given to participants and screenshots?
    \answerNA{NA}
  \item Did you describe any potential participant risks, with mentions of Institutional Review Board (IRB) approvals?
    \answerNA{NA}
  \item Did you include the estimated hourly wage paid to participants and the total amount spent on participant compensation?
    \answerNA{NA}
   \item Did you discuss how data is stored, shared, and deidentified?
   \answerYes{Yes} 
   
\end{enumerate}

\end{enumerate}

\onecolumn

\appendix

\section*{Appendix}

\begin{table*}[ht!]
    \caption{Fields in the Discord Server Object. These fields provide metadata for each server included in the dataset.}
    \centering
    \renewcommand{\arraystretch}{1.3} % Espaçamento vertical
    \setlength{\tabcolsep}{5pt} % Espaçamento horizontal
    \begin{tabular}{|p{3.5cm}|p{3.5cm}|p{8cm}|}
        \hline
        \textbf{Field} & \textbf{Type} & \textbf{Description} \\ 
        \hline
        slug & string & A unique identifier combining the server name and ID, used as a shorthand or URL-friendly reference. \\ 
        \hline
        id & string & A unique numerical identifier for the server. \\ 
        \hline
        name & string & The human-readable name of the server. \\ 
        \hline
        description & string & A brief description or identifier for the server. \\ 
        \hline
        icon & string & URL to the server's icon image. \\ 
        \hline
        splash & string & URL to the splash image, typically used for larger visual displays or promotional purposes. \\ 
        \hline
        banner & string & Identifier for the server's banner image, used for branding or visual purposes. \\ 
        \hline
        approximate\_presence\_ count & number & The estimated number of currently active users on the server. \\ 
        \hline
        approximate\_member\_ count & number & The estimated total number of members in the server. \\ 
        \hline
        premium\_subscription\_ count & number & The number of members with premium subscriptions (e.g., Discord Nitro) on the server. \\ 
        \hline
        preferred\_locale & string & The server's preferred language, represented by its locale code. \\ 
        \hline
        auto\_removed & boolean & Indicates whether the server was automatically removed from a directory or platform. \\ 
        \hline
        discovery\_splash & string & An identifier for the discovery splash image, used for promoting the server in public directories. \\ 
        \hline
        primary\_category\_id & number & ID representing the primary category or classification of the server. \\ 
        \hline
        vanity\_url\_code & string & Custom short URL code for the server. \\ 
        \hline
        is\_published & boolean & Indicates whether the server is published and discoverable in public directories. \\ 
        \hline
        keywords & array of strings & A list of keywords associated with the server for discoverability. \\ 
        \hline
        features & array of strings & A list of special features enabled for the server, providing enhanced functionality or customization. \\ 
        \hline
        created\_at & string (ISO 8601) & The timestamp indicating when the server was created. \\ 
        \hline
        reasons\_to\_join & array of strings & A list of promotional highlights or unique selling points of the server. \\ 
        \hline
        social\_links & array of strings & Links to external social media or community pages associated with the server. \\ 
        \hline
        about & string & A detailed description of the server's purpose, history, and features. \\ 
        \hline
        category\_ids & array of numbers & A list of IDs representing the categories the server is associated with. \\ 
        \hline
    \end{tabular}
    \label{tab:discord_server_fields}
\end{table*}

\begin{table*}[ht!]
    \caption{Fields in the Discord Message Object. Fields marked with '?' are optional. References [1], [2], etc., refer to detailed specifications in the Discord API documentation.}
    \centering
    \renewcommand{\arraystretch}{1.3
    } % Espaçamento vertical
    \setlength{\tabcolsep}{5pt} % Espaçamento horizontal
    \begin{tabular}{|p{3.5cm}|p{3.5cm}|p{8cm}|}
        \hline
        \textbf{Field} & \textbf{Type} & \textbf{Description} \\ 
        \hline
        id & snowflake & ID of the message \\ 
        \hline
        channel\_id & snowflake & ID of the channel the message was sent in \\ 
        \hline
        author & user object & The author of this message (not guaranteed to be a valid user, see below) \\ 
        \hline
        content & string & Contents of the message \\ 
        \hline
        timestamp & ISO8601 timestamp & When this message was sent \\ 
        \hline
        edited\_timestamp & ?ISO8601 timestamp & When this message was edited (or null if never) \\ 
        \hline
        tts & boolean & Whether this was a TTS message \\ 
        \hline
        mention\_everyone & boolean & Whether this message mentions everyone \\ 
        \hline
        mentions & array of user objects & Users specifically mentioned in the message \\ 
        \hline
        mention\_roles & array of role object IDs & Roles specifically mentioned in this message \\ 
        \hline
        mention\_channels? & array of channel mention objects & Channels specifically mentioned in this message \\ 
        \hline
        attachments & array of attachment objects & Any attached files \\ 
        \hline
        embeds & array of embed objects & Any embedded content \\ 
        \hline
        reactions? & array of reaction objects & Reactions to the message \\ 
        \hline
        nonce? & integer or string & Used for validating a message was sent \\ 
        \hline
        pinned & boolean & Whether this message is pinned \\ 
        \hline
        webhook\_id? & snowflake & If the message is generated by a webhook, this is the webhook's ID \\ 
        \hline
        type & integer & Type of message \\ 
        \hline
        activity? & message activity object & Sent with Rich Presence-related chat embeds \\ 
        \hline
        application? & partial application object & Sent with Rich Presence-related chat embeds \\ 
        \hline
        application\_id? & snowflake & If the message is an Interaction or application-owned webhook, this is the ID of the application \\ 
        \hline
        flags? & integer & Message flags combined as a bitfield \\ 
        \hline
        message\_reference? & message reference object & Data showing the source of a crosspost, channel follow add, pin, or reply message \\ 
        \hline
        referenced\_message? [4] & ?message object & The message associated with the message\_reference \\ 
        \hline
        thread? & channel object & The thread that was started from this message, includes thread member object \\ 
        \hline
        components? & array of message components & Sent if the message contains components like buttons, action rows, or other interactive components \\ 
        \hline
        sticker\_items? & array of message sticker item objects & Sent if the message contains stickers \\ 
        \hline
        stickers? & array of sticker objects & Deprecated: the stickers sent with the message \\ 
        \hline
        position? & integer & A generally increasing integer (there may be gaps or duplicates) that represents the approximate position of the message in a thread \\ 
        \hline
        role\_subscription\_data? & role subscription data object & Data of the role subscription purchase or renewal that prompted this ROLE\_SUBSCRIPTION\_PURCHASE message \\ 
        \hline
        resolved? & resolved data & Data for users, members, channels, and roles in the message's auto-populated select menus \\ 
        \hline
        poll? [2] & poll object & A poll! \\ 
        \hline
        call? & message call object & The call associated with the message \\ 
        \hline
    \end{tabular}
    \label{tab:discord_message_fields}
\end{table*}

\twocolumn

\end{document}